\documentclass[useAMS,usenatbib,referee]{mn2e}
\usepackage{graphicx}
\usepackage{natbib}
\usepackage{subfigure}

\usepackage{subfigure}
\bibpunct{(}{)}{;}{a}{}{,}
\title{H$_2$ reformation in post-shock regions}

\newcommand{\aap}{A\&A}
\newcommand{\apj}{ApJ}
\newcommand{\jcp}{J.~Chem.~Phys.}
\newcommand{\mnras}{MNRAS}
\newcommand{\apjs}{ApJS}

\author[H.~M.~Cuppen, L.E.~Kristensen and E.~Gavardi]{H.~M.~Cuppen, L.E.~Kristensen and E.~Gavardi\\
Leiden Observatory, Leiden University, P.O. Box 9513, 2300 RA Leiden, The Netherlands
}

\begin{document}
\maketitle

\begin{abstract}
H$_2$ formation is an important process in post-shock regions, since H$_2$ is an active participant in the cooling and shielding of the environment. The onset of H$_2$ formation therefore has a strong effect on the temperature and chemical evolution in the post shock regions. We recently developed a model for H$_2$ formation on a graphite surface in warm conditions. The graphite surface acts as a model system for grains containing large areas of polycyclic aromatic hydrocarbon structures. Here this model is used to obtain a new description of the H$_2$ formation rate as a function of gas temperature that can be implemented in molecular shock models. The H$_2$ formation rate is substantially higher at high gas temperatures as compared to the original implementation of this rate in shock models, because of the introduction of H atoms which are chemically bonded to the grain (chemisorption).
Since H$_2$ plays such a key role in the cooling, the increased rate is found to have a substantial effect on the predicted line fluxes of an important coolant in dissociative shocks  [O~\textsc{I}] at 63.2 and 145.5~$\mu$m. With the new model a better agreement between model and observations is obtained.  
Since one of the goals of Herschel/PACS will be to observe these lines with higher spatial resolution and sensitivity than the former observations by ISO-LWS, this more accurate model is very timely to help with the interpretation of these future results.
\end{abstract}


\section{Introduction}
H$_2$ formation is an important process in post-shock regions, since H$_2$ is an active participant in the cooling and shielding of the environment. The onset of H$_2$ formation therefore has a strong effect on the temperature and chemical evolution in the post shock regions. 
This letter aims to re-evaluate the estimations and assumptions concerning H$_2$ reformation in molecular shock models. We mainly focus on jump-type (J-type) shocks. These can reach velocities of several hundred km s$^{-2}$ causing an increase of the gas temperature $>10^5$~K. This will, in turn, result in the dissociation of molecular hydrogen, in contrast with continuous-type (C-type) shocks where H$_2$ stays intact. The dissociation of H$_2$ leads to the first cooling of the region \citep{Kwan:1977, London:1977} to roughly 4300~K. Since all molecular material that is responsible for cooling under normal circumstances (and in C-type shocks), is dissociated, O cooling is the dominant cooling mechanism \citep{Flower:2003}. At some point, H$_2$ starts to reform again on the surfaces of dust grains. Because of its high abundance, H$_2$ cooling is more efficient than O cooling and it cools the gas phase until roughly 100~K when cooling by H$_2$O takes over. O emission is classically used as a shock tracer, since it requires H$_2$ dissociation and high temperatures, representative for J-type shocks \citep[$E_{\rm up} \sim 230-330$ K; \emph{e.g.},][]{Nisini:2002}. Observations of O emission are typically complemented by H$_2$ observations, which gives more lines with energy levels spread out over $\sim$ 500--15\,000 K. The combination of both observables can, in principle, provide a good handle on the physical conditions in shocks. However, accurate predictions and models are needed to translate these observations to physical conditions.

Detailed molecular shock models can help us with this translation. They provide predictions of observables depending on a set of input parameters describing the physical conditions. Currently, most shock models include a number of ``educated'' estimations for parameter values. This work aims at minimising the uncertainty on at least one of these parameters: the reformation rate of H$_2$ on grain surfaces. As explained above, the exact conditions under which H$_2$  reforms is an important component in the evolutionary sequence and therefore in the interpretation of the observations. Often in gas phase models, the efficiency, \emph{i.e.}, the fraction of H atoms hitting a grain that leaves the surface as H$_2$, is assumed to be unity and independent of temperature and pressure. Some further assumptions are made on the sticking of hydrogen atoms on the grains.

Grains can be of silicate or carbonaceous nature and they can be covered by a mantle of ice. Here we focus on bare grains, since processing of ices in shocks, like inertial sputtering, has probably lead to the evaporation of the ice mantles. The chemical nature of the grain surface has an effect on the H$_2$ formation rate in terms of both sticking and formation efficiency. 
We recently developed a model for molecular hydrogen formation on a graphite surface in warm conditions, such as shocks, PDRs and regions around AGNs. The graphite surface acts as a model system for grains containing large areas of polycyclic aromatic hydrocarbon (PAH) structures \citep{Pendleton:2002}. This type of dust  takes up a large fraction of the interstellar dust surface area and its interaction with hydrogen has been studied substantially in laboratory and theoretical studies. Our kinetic Monte Carlo model was tested against four independent experimental observations (H$_2$ abstraction cross section, saturation coverage, surface structures, and desorption behaviour) and a good agreement was obtained \citep{Cuppen:2008,Gavardi:2009}. The present letter extends this model to shock conditions by determining the H$_2$ formation efficiencies under these conditions. The improved rate is then introduced into a shock code to study its effect on the temperature and chemistry. The line fluxes of the [O~\textsc{I}] at 63.2 and 145.5~$\mu$m are found to be strongly affected by the new H$_2$ formation. In a comparative study between model and observations for HH54, \cite{Giannini:2006} found an overestimation of the 63.2~$\mu$m emission by at least a factor of 50. Here we will compare observations of HH46  to our new model and a much better agreement is obtained.

\section{The Monte Carlo method}
For a detailed description of method and its parametrisation, we refer to \cite{Cuppen:2008} and \cite{Gavardi:2009}. In short, the model uses a continuous-time, random-walk Monte Carlo scheme \citep{Chang:2005} where hydrogen atoms can physisorb or chemisorb to the substrate which is represented by a honeycomb structure mimicking the graphite surface. The atoms can move across the surface and desorb from the surface. Physisorbed atoms can become chemisorbed after crossing a small barrier. The barriers and binding energies are obtained from theoretical methods. In accordance with experiments \citep{Hornekaer:2006I}, hydrogen atoms can cluster together on the surface. Some dimer structures, two hydrogen atoms chemisorbed to carbon atoms in close vicinity of each other, are found to be more stable than two, well-separated, monomers. The same holds for more complex structures. The model takes structures up to tetramers explicitly into account.

Physisorbed atoms are found to play a key role in the formation of these complex structures, even close to room temperature, where the lifetime of physisorbed atoms on the surface is short. Since their diffusion rate is high \citep{Bonfanti:2007}, they scan a large part of the surface, find a chemisorbed H atom, and can form a dimer structure. 

\cite{Cuppen:2008} describe a comparison between the model results and H$_2$ abstraction experiments from \cite{Zecho:2002}. In order to obtain agreement, a mechanism was introduced where the excess energy of the chemisorbing H atom does not dissipate instantaneously, but can be used to cross, for instance, the H$_2$ formation barrier.

The Monte Carlo algorithm starts by the deposition of the first atom on the surface. The deposition rate is determined by the gas phase abundance of atomic hydrogen, $n({\rm H})$, the sticking fraction to the grain, $S$, and the velocity of atoms in the gas phase
\begin{equation}
v_{\rm H} = \sqrt{\frac{8kT_{\rm gas}}{\pi m_{\rm H}}}
\end{equation}
with $k$ the Boltzmann constant, $T_{\rm gas}$ the gas phase temperature of the incoming atom, and $m_{\rm H}$ the mass of an H atom.
Then all possible events and the corresponding rates, $R_i$, of the deposited atom are considered depending on its environment and binding state. The time at which the atom will undergo an event is determined by
\begin{equation}
t_{\rm event} = t_{\rm current} - \frac{\ln\left(X\right)}{\sum_i R_i}
\end{equation}
with $t_{\rm current}$ the current time, at which the first atom is deposited, and $X$ a random number between 0 and 1. This will be compared against the time of the second deposition. The first event will happen and new possible events are evaluated. This cycle is repeated until the end of the simulation.
Event rates are determined by their barriers $E_i$
\begin{equation}
R_i = \nu\exp\left(-\frac{E_i}{T_{\rm grain}}\right)
\end{equation}
with $T_{\rm grain}$ the surface temperature of the grain and $\nu$ the attempt frequency ($10^{13}$~Hz). An overview of all barriers that are included in the model is given in \cite{Cuppen:2008}. We have increased the physisorption desorption energy from 500 to 580~K to mimic the increased binding energies at step edges and other defects. 

\begin{figure}
\begin{center}
\includegraphics[width=0.4\textwidth]{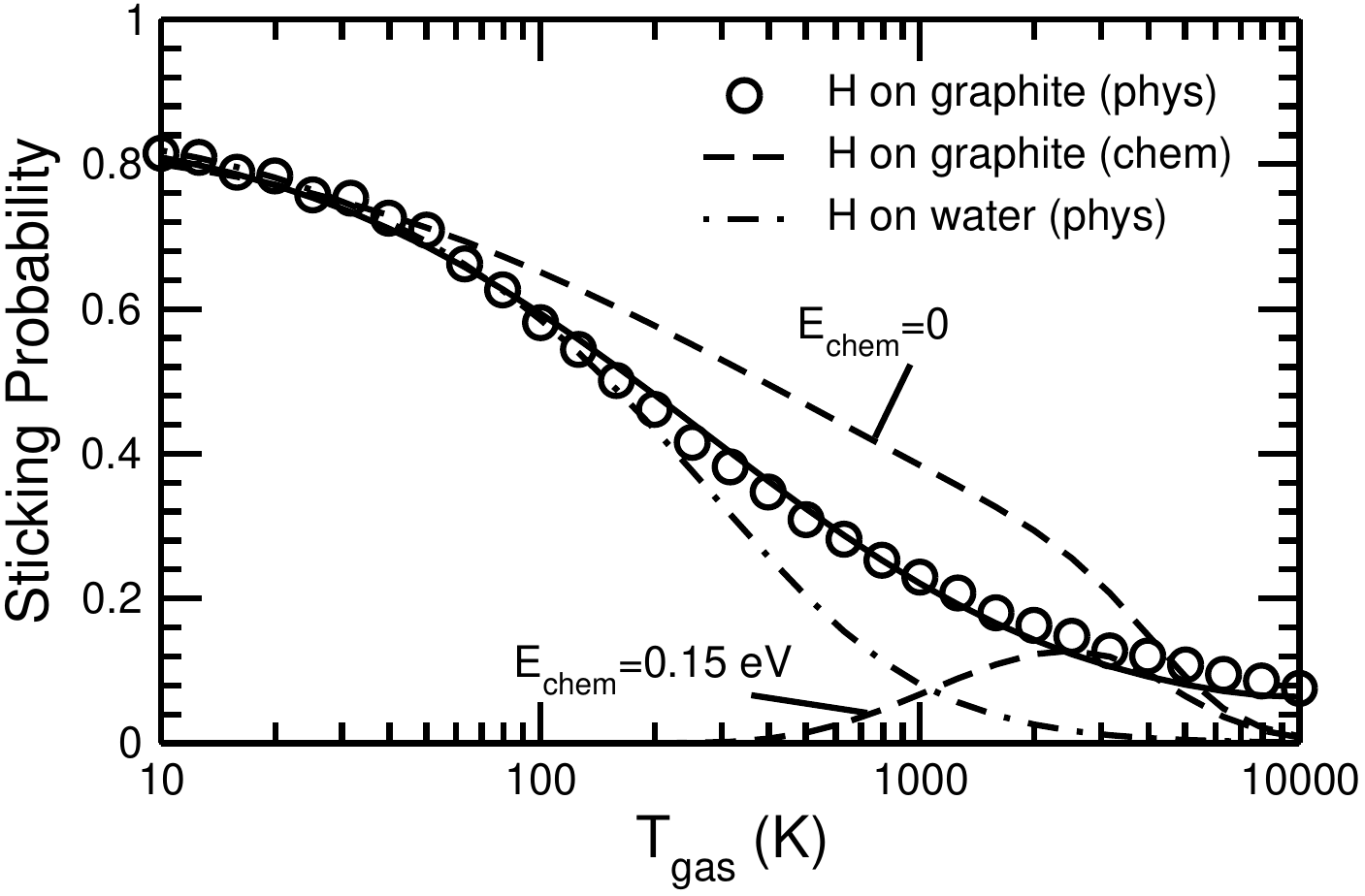}
\end{center}
\caption{The sticking probability on graphite for physisorbed (circles and solid curve) and chemisorbed (dashed curves) H-atoms with different chemisorption barriers. The sticking probability of H-atoms to water ice is plotted for comparison (Burke \& Hollenbach 1983).}
\label{stick}
\end{figure}

\section{Sticking probability}
The original implementation of the H$_2$ formation rate in the model of \cite{Flower:1985} assumes the efficiency to be unity independent of temperature and pressure. For the sticking probability, an expression describing the sticking of H atoms to water ice is used \citep{Burke:1983}.
This expression is plotted as function of gas temperature in Fig.~\ref{stick} (dash-dotted curve).  However, grains are most likely not covered with water ice in post-shock regions. Inertial sputtering efficiently removes the ice mantles even for low velocity shocks \citep{Guillet:2009II}.  

Hydrogen atoms stick to graphite with a different efficiency at high gas phase temperature than to water. As explained earlier, hydrogen atoms can either be physisorbed or chemisorbed to the surface with a barrier between both potential wells, depending on the local environment. The trapping to both states has different probabilities and different dependencies on gas temperature. \cite{Sha:2005} performed quantum studies on the trapping of H atoms on graphite surface. They present the sticking probability after 1.0~ps as a function of incident energy. Here only the chemisorption channel was taken into account, and physisorption was ignored. To obtain a generalised expression that can be included in the Monte Carlo model, their results are described in terms of the product of a sticking probability and probability to cross the chemisorption barrier
\begin{equation}
S_{\rm chem} = \frac{\exp\left(-\frac{E_{\rm chem}}{T_{\rm gas}}\right)}{1+5\times10^{-2}\sqrt{T_{\rm{gas}} + T_{{\rm grain}}}+ 1\times10^{-14}T_{\rm{gas}}^4}.
\end{equation} 
with temperatures in K. 
Figure \ref{stick} plots this chemisorption sticking probability with the dashed curves for $E_{\rm chem} = 0$ and $E_{\rm chem} = 0.15$~eV. The first corresponds to sticking in a para-dimer configuration; the latter to sticking of monomers and corresponds to the work by \cite{Sha:2005}.

\begin{figure}
\begin{center}
\includegraphics[width=0.39\textwidth]{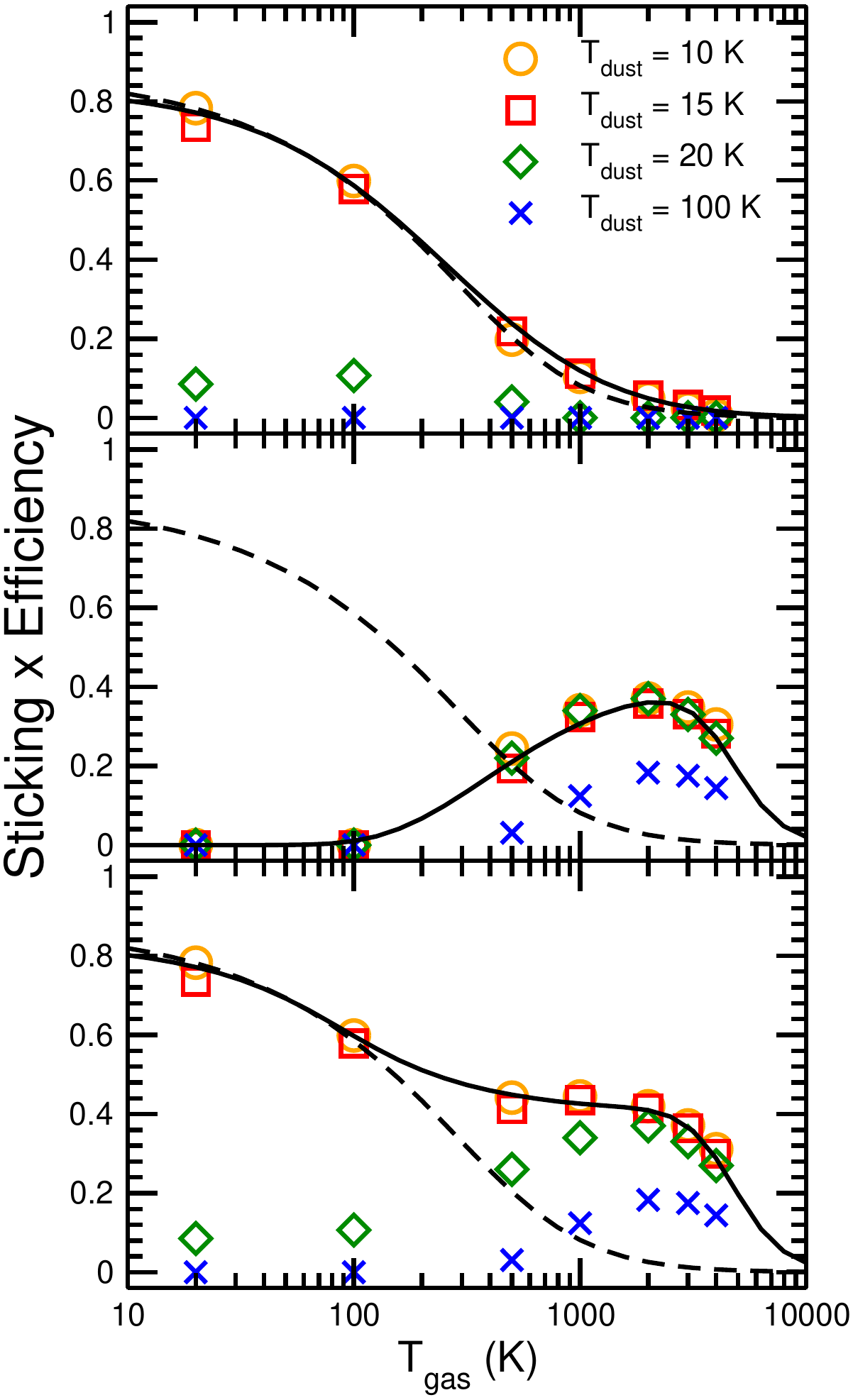}
\end{center}
\caption{The combined H-atom sticking and H$_2$ formation efficiency as a function of gas phase temperature for four different grain temperatures. The contributions of physisorbed (top panel) and chemisorbed (middle panel) atoms to H$_2$ formation are separated. The bottom panel gives the total of $S\times\eta$. The black dashed curve gives $S\times\eta$ which was originally implemented in the shock model.}
\label{eff}
\end{figure}

The remaining non-sticking atoms still have an opportunity to become trapped in the physisorption well. To obtain this sticking probability we follow \cite{Burke:1983}, who used the so-called soft-cube method \citep{Logan:1968}. An incoming gas phase hydrogen atom is followed according to
\begin{equation}
\frac{{\rm d}^2 z_{\rm gas}}{{\rm d}t^2} = \frac{F}{m_{\rm H}}
\end{equation}
with $z_{\rm gas}$ the $z$-coordinate of the gas phase atom (normal to the surface) and $F$ the force acting on this particle. The approached grain atom moves according to 
\begin{equation}
\frac{{\rm d}^2 z_{\rm grain}}{{\rm d}t^2} = -\frac{F}{m_{\rm grain}} - \omega^2 z_{\rm grain}
\end{equation}
with $\omega = k \theta_D/\hbar$ 
where $\theta_D$ is the Debye temperature of the solid and $\hbar$ Planck's constant divided by $2\pi$. The gas phase atom starts from a large distance with a velocity following a Maxwell-Boltzmann distribution and the grain atom oscillates starting with a random phase and an amplitude of
\begin{equation}
z_{\rm grain, max} = \sqrt{\frac{2kT_{\rm grain}}{\omega^2m_{\rm grain}}}.
\end{equation} 
This method is found to give a reasonable agreement with experimental sticking probabilities of inert gas phase particles sticking to metal surfaces and we can therefore use it for the non-reactive trapping of hydrogen to graphite (physisorption). 
To determine the force, $F$, we use an accurate H-coronene physisorption potential, which was fitted to \emph{ab initio} data on MP2 level \citep{Bonfanti:2007} and which accurately reproduces experimental desorption data of H physisorbed to graphite. The sticking probability that is determined in this way is plotted by the open circles in Fig.~\ref{stick}. The solid line represents a fit to the data:
\begin{equation}
S_{\rm phys}=\left(1+a_1\sqrt{T_{\rm{gas}} + T_{{\rm grain}}}+ a_2T_{\rm{gas}} - a_3T_{\rm{gas}}^2\right)^{-1},
\label{S_phys}
\end{equation} 
with $a_1=4.2\times10^{-2}$~K$^{-1/2}$, $a_2=2.3\times10^{-3}$~K$^{-1}$, and $a_3=1.3\times10^{-7}$~K$^{-2}$.

\section{H$_2$ formation efficiency}
The combination of the sticking probability and the H$_2$ formation efficiency, $\eta$, is determined by Monte Carlo simulations. Simulation runs for a grid of different gas temperatures, H-atom gas abundances, and surface temperatures are performed. The shock model assumes the surface temperature to remain at a constant value of 15~K. The main heating mechanism for grains in shock regions is through UV photons. For the density and velocity regime that is covered here hardly any FUV photons are produced in the dissociation of the gas and the dust remains cold \citep{Hollenbach:1989}. Therefore simulations are performed at a constant low grain temperature, but with varying gas temperature, which affects the initial sticking to the surface. 

After a short initial equilibration time, a constant H$_2$ formation efficiency is reached. This equilibration time scales with $n({\rm H})$. For $n({\rm H})= 10^6$ cm$^{-3}$, it takes roughly $5 \times 10^{-3}$~years to reach steady state. In comparison, $n({\rm H})$ in the post-shock gas is between $2\times 10^4 - 2\times 10^7$ cm$^{-3}$ and the onset for H$_2$ reformation occurs at longer timescales than the $n(H)$-dependent equilibration time ($> 2\times$). Figure \ref{eff} shows the final $S\times\eta$ as a function of gas temperature for different surface temperatures. The final values are obtained by averaging the steady state values obtained in five independent Monte Carlo simulation runs, with different initial seeds. Very little variation between the individual runs was found. Little to no dependence on $n({\rm H})$ was observed over the relevant range.

As explained earlier, H$_2$ can form from chemisorbed or physisorbed atoms. In Fig.~\ref{eff}, we distinguish between both mechanisms. The top panel shows the contribution to H$_2$ formation by physisorbed atoms, the middle panel by chemisorbed atoms, and the bottom panel plots the sum of both contributions. 
The physisorbed mechanism is dominant at low gas and surface temperatures. At high gas temperatures most of hydrogen atoms scatter from the grain and for high grain temperatures the residence time of the atoms becomes shorter than the time between two depositions. Chemisorption requires the crossing of a barrier and is only efficient at higher temperatures. The dashed curves in Fig.~\ref{eff} represents the original implementation of the product of the sticking probability and the efficiency ($=1$), resembling the physisorption contribution. Above $T_{\rm gas} =500$~K, the old and the new results start to deviate because of the introduction of the chemisorption mechanism in the new model.

The solid curves in Fig.~\ref{eff} give fits to the simulation data. These descriptions can be included in shock models to accurately describe the H$_2$ formation rate. We find 
\begin{equation}
S_{\rm phys}\eta_{\rm phys}=\frac{0.8}{1+ 4\times 10^{-3}T_{\rm gas} +  2\times 10^{-6}T_{\rm gas}^2}
\label{eta_phys}
\end{equation} 
for the physisorption contribution for $T_{\rm grain} < 20$ K and
\begin{equation}
S_{\rm chem}\eta_{\rm chem}=\frac{0.45\exp(-\frac{380}{T_{\rm gas}})}{1+ 2\times 10^{-15}T_{\rm gas}^4}
\label{eta_chem}
\end{equation} 
for the chemisorption contribution. These expressions are valid for $2\times 10^4 \leq n({\rm H}) \leq 2\times 10^7$ cm$^{-3}$.

\section{Shock model}

\subsection{Model description and updates}
To test the impact of the above results, the molecular shock model described in \citet{Guillet:2009II} has been chosen as an application. This model has been chosen since it is a sophisticated 1D model in which chemistry and cooling is treated in parallel with the magneto hydrodynamics. It is capable of treating both C-type and J-type shocks in a self-consistent manner. Grains are included as well as sputtering of the grain mantles. For the purposes of this paper only dissociative J-type shocks are considered, since H$_2$ reformation is not at play in C-type shocks.
The H$_2$ reformation scheme has been adopted as follows:
\begin{equation}
R_{\rm old}({\rm H}_2) = \eta f({\rm H}) n({\rm grain}) \pi r^2_{\rm grain}S_{\rm water}
\end{equation}
where $f$(H) the H-atom flux, $n$(grain) the number of grains, and $\pi r^2_{\rm grain}$ the geometric grain cross section. The efficiency is always assumed to be equal unity, and the sticking coefficient is given by the dash-dotted curve in Fig.~\ref{stick}.
The model has been updated to include the new results for H$_2$ reformation
\begin{equation}
R_{\rm new}({\rm H}_2) =  f({\rm H}) n({\rm grain}) \pi r^2_{\rm grain}(S_{\rm phys}\eta_{\rm phys}+S_{\rm chem}\eta_{\rm chem})
\end{equation}
using Eqs.~\ref{eta_phys} and \ref{eta_chem}. 

\setlength{\tabcolsep}{2pt}

\begin{table}
\caption{Predicted H$_2$ 1--0 $S$(1) and [O~\textsc{I}] 63.2 $\mu$m and 145.5 $\mu$m line fluxes in the old and new version.}
\label{O I}
\begin{tabular}{ccllllllll}
\hline
$n_{\rm H}$ & $v$ & \multicolumn{2}{c}{H$_2$ 1--0 $S$(1)} & \multicolumn{2}{c}{[O~\textsc{I}] 63.2 $\mu$m} & \multicolumn{2}{c}{[O~\textsc{I}] 145.5 $\mu$m} \\
(cm$^{-3}$) & (km/s) & \multicolumn{2}{c}{(W\,m$^{-2}$\,sr$^{-1}$)}  &\multicolumn{2}{c}{(W\,m$^{-2}$\,sr$^{-1}$)}  & \multicolumn{2}{c}{(W\,m$^{-2}$\,sr$^{-1}$)}\\
 &  & \multicolumn{1}{c}{old} & \multicolumn{1}{c}{new} & \multicolumn{1}{c}{old} & \multicolumn{1}{c}{new}& \multicolumn{1}{c}{old} & \multicolumn{1}{c}{new} \\
\hline
10$^3$ & 30  &2.97(-8) &2.97(-8)& 2.97(-9) & 2.32(-9) & 1.18(-10) & 0.84(-10) \\
       & 40  &1.66(-8) &2.40(-8)& 6.76(-7) & 4.42(-7) & 2.38(-8) & 1.46(-8)   \\
       & 50  &1.43(-8) &1.72(-8)& 6.47(-7) & 4.23(-7) & 2.08(-8) & 1.27(-8)   \\
10$^4$ & 30  &8.72(-8) &1.55(-7)& 3.00(-6) & 1.76(-6) & 8.44(-8) & 4.89(-8)   \\
       & 40  &1.07(-7) &1.40(-7)& 1.57(-6) & 0.62(-6) & 4.12(-8) & 1.49(-8)   \\
       & 50  &1.13(-7) &9.48(-8)& 1.18(-6) & 0.45(-6) & 2.96(-8) & 1.00(-8)   \\
10$^5$ & 30  &3.74(-7) &5.88(-7)& 7.16(-6) & 4.03(-6) & 1.94(-7) & 1.10(-7)   \\
       & 40  &4.06(-7) &5.67(-7)& 3.41(-6) & 0.95(-6) & 9.07(-8) & 2.49(-8)   \\
       & 50  &4.02(-7) &2.47(-7)& 2.08(-6) & 0.42(-6) & 5.41(-8) & 0.95(-8)   \\
\hline
\end{tabular}\\
$a(-b)$ refers to $a \times 10^{-b}$
\end{table}

\subsection{Results}
\begin{figure}
\begin{center}
\includegraphics[width=0.49\textwidth]{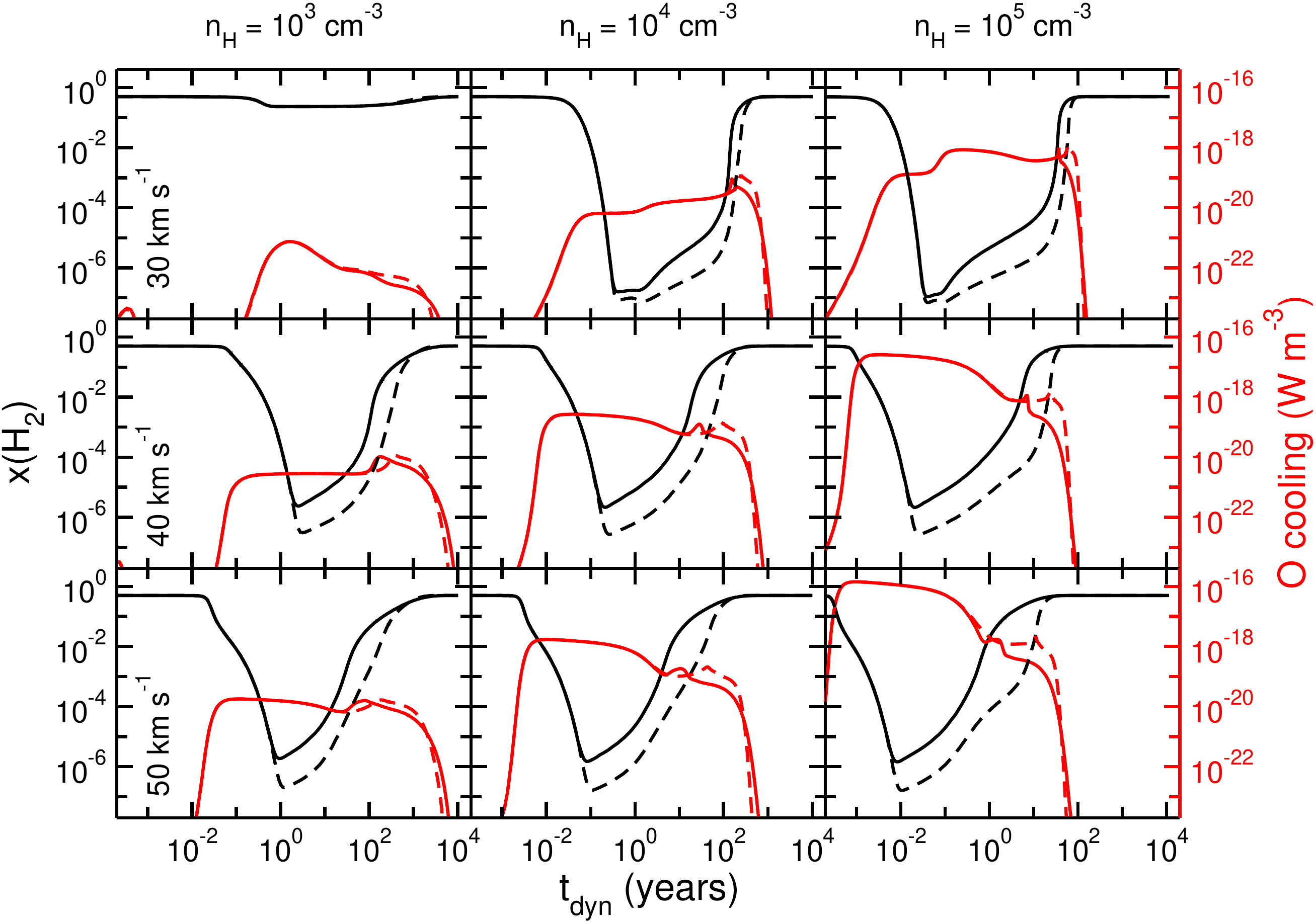}
\end{center}
\caption{The relative abundance of H$_2$ (left axis in black) and the O cooling (right axis in orange) for a grid of different velocities (30, 40, and 50~km s$^{-1}$) and initial densities (10$^3$, 10$^4$, and 10$^5$~cm$^{-3}$). The old model results are represented by the dashed lines, the new by the solid lines.}
\label{multi}
\end{figure}

Figure \ref{multi} plots the relative H$_2$ abundance and the O cooling contribution for a grid of different shock simulations using different initial conditions in terms of velocity and density. The figure clearly shows that H$_2$ is reformed at earlier times, especially for high densities and velocities. H$_2$ takes therefore over as the main coolant earlier and at higher temperatures than previously seen. In general, this results in higher emission in high-$v$, high-$J$ lines of H$_2$ at the cost of emission in low-$J$ lines. The changes in $v$=1--0 $S$(1) emission, for instance, are small and sensitive to the initial physical conditions as is shown in Table \ref{O I}. Changes within a factor of 1.8 are obtained, in both directions. The total H$_2$ cooling summed over all lines is  affected by less than 8\%.
Since H$_2$ takes over as the main coolant earlier than in the older models, the timescale for, for example, oxygen cooling is shorter. This has the effect that the two [O~\textsc{I}] lines at 63.2 and 145.5 $\mu$m have been over-estimated by factors of 1.5--5 in the older models for the same conditions as is indicated in Table~\ref{O I}. 

As discussed above, we assume that the grain temperature does not change throughout the shock. Even if the dust would be heated to temperatures above 20 K, this will be at early times when H$_2$ reformation occurs mainly through the chemisorption mechanism which is less temperature dependent. By the time that the physisorption mechanism starts to dominate, the dust has probably cooled down to lower temperatures. A grid of models with $S_{\rm phys}\eta_{\rm phys}=0$ shows indeed that the H$_2$ production is only affected at late times. The [O \textsc{I}] fluxes are a factor of 0.33--0.87 lower, mainly because the cold post-shock gas is not fully converted to H$_2$.

\cite{Nisini:2002} have measured the far-infrared spectra (45--197~$\mu$m) of 28 low-luminosity young embedded objects using the Infrared Space Observatory. Their sample includes the detection of ten 63.2~$\mu$m [O~\textsc{I}]  and six 145.5~$\mu$m [O~\textsc{I}] lines. HH46 IRS is a typical source among these and we will use this for observational comparison. Assuming uniform emission, the observed emission of $(5.25 \pm 0.29) \times 10^{-15}$~W\,m$^{-2}$ of the 63.2~$\mu$m line corresponds to $(3.49\pm 0.19) \times 10^{-8}$~W\,m$^{-2}$\,sr$^{-1}$ for the 80\arcsec\ aperture of the LWS receiver of ISO, the emission of the 145.5~$\mu$m line corresponds to $(1.9\pm0.5) \times 10^{-9}$~W\,m$^{-2}$\,sr$^{-1}$. These values can be directly compared to the values in Table~\ref{O I}. With the old implementation of the H$_2$ formation rate, almost all dissociative shocks ($v > 25$~km\,s$^{-1}$) overproduce the [O~\textsc{I}] 63.2~$\mu$m and 145.5~$\mu$m emission with respect to the observed flux by at least factors of 18 and 11, respectively ($v = 50$~km\,s$^{-1}$ and $n_{\rm H} = 10^{3}$~cm$^{-3}$). For higher densities  discrepancies increase rapidly. For the new model, the predicted [O~\textsc{I}] line fluxes are smaller and the density dependence is also reduced. The observed values are approached by factors of 12 and 5, respectively, with very little dependence on the density. High velocities appear to be required to approach the observed values. In general, the agreement of the new models with the observations is better than for the old model, especially considering that the best agreement in the old model is only obtained for very specific conditions whereas in the new model a larger range of initial densities and velocities produces a good agreement.
\cite{GarciaLopez:2010} reported the H$_2$ 1--0 $S$(1) line flux in six different knots of HH46. These observations are performed with a much smaller beam (6 times 0\arcsec3) than the ISO observations. The total flux corresponds to $(8.4\pm0.1) \times 10^{-7}$~W\,m$^{-2}$\,sr$^{-1}$ which approached within a factor of 3.4 by the models for $v = 50$~km\,s$^{-1}$ and $n_{\rm H} = 10^{5}$~cm$^{-3}$ (best fit for [O~\textsc{I}]). As mentioned before, the H$_2$ 1--0 $S$(1) flux is rather sensitive to the physical conditions. For this particular combination of conditions the agreement with the model is less than with the old model. For lower initial velocity an improved agreement between model and observations is obtained.

\section{Conclusions}
To summarise, we have introduced a new model for the H$_2$ formation rate on carbonaceous surfaces as a function of gas temperature at a constant low grain temperature.  This is obtained by a state-of-the-art model of H absorption on a graphite surface which includes both physisorption and chemisorption of complex structures. At low gas temperatures, physisorbed atoms can stick to the surface whereas chemisorbed atoms do not have enough energy to overcome the barrier for sticking. The H$_2$  formation rate is dominated by recombination of physisorbed atoms in this regime. At high gas temperatures, most incoming atoms either scatter from the surface or become trapped in a chemisorption well. The H$_2$ formation rate is therefore mainly determined by chemisorbed atoms. This new model for the H$_2$ formation rate is used as input for a molecular shock model. The rate is substantially higher at high gas temperatures as compared to the original implementation of this rate in shock models. This increased rate has a substantial effect on the predicted line fluxes of [O~\textsc{I}] at 63.2 and 145.5~$\mu$m and a better agreement between model and observations is obtained than with the previous model.  
One of the goals of Herschel/PACS will be to observe these lines with higher spatial resolution and sensitivity than the former observations by ISO-LWS.  This more accurate model will help with the interpretation of these future results.

\section*{acknowledgements}
EG would like to thank the Lifelong Learning Program from the European Union for her stay at the Leiden University. HC is supported by the Netherlands Organization for Scientific Research (NWO).

\end{document}